\documentclass{ws-ijbc}
\usepackage{ws-rotating}     
\usepackage{graphicx}
\usepackage{epstopdf}
\begin{document}
\catchline{}{}{}{}{} % Publisher's Area please ignore

\markboth{Katsanikas and Wiggins}{Phase Space Analysis of the Non-Existence of Dynamical Matching  in a  Stretched  Caldera Potential Energy Surface}
\title{Phase Space Analysis of the Non-Existence of Dynamical Matching  in a  Stretched  Caldera Potential Energy Surface}
\author{MATTHAIOS  KATSANIKAS  AND  STEPHEN  WIGGINS}
\address{School of Mathematics,  University of Bristol, University Walk, Bristol BS8 1TW, UK \\ matthaios.katsanikas@bristol.ac.uk, s.wiggins@bristol.ac.uk}


\maketitle

\begin{history}
\received{(to be inserted by publisher)}
\end{history}

\maketitle
\begin{abstract}
In this paper we continue our studies of the two dimensional caldera potential energy surface in a parametrized family that allows for a study of the effect of symmetry on the
phase space structures that govern how trajectories enter, cross, and exit the region of the caldera. As a particular form of trajectory crossing, we are able to determine the effect of symmetry and phase space structure on dynamical matching. We show that there is a critical value of the symmetry parameter which controls the phase space structures responsible for the manner of crossing, interacting with the central region (including trapping in this region) and exiting the caldera. We provide an explanation for the existence of this critical value in terms of the behavior of the H{\'e}non stability parameter for the associated periodic orbits.
\end{abstract}


\section{Introduction}
\label{intro}
The caldera potential energy surface arises in many organic chemical reactions, such as   the  vinylcyclopropane-cyclopentene rearrangement \cite{baldwin2003,gold1988}, the stereomutation of cyclopropane \cite{doubleday1997}, the degenerate rearrangement  of bicyclo[3.1.0]hex-2-ene\cite{doubleday1999,doubleday2006} or 5-methylenebicyclo[2.1.0]pentane \cite{reyes2002}. The name ‘’caldera’’ arises from the fact that the shape of this potential energy surface resembles that of an erupted volcano.  The central minimum of the potential marks a flat region of the potential energy surface that plays a crucial influence on the main four routes of trajectories crossing this region. The beginning and end of these four routes correspond   to the four regions influenced by the index-1 saddles. These regions are the only locations for allowing the molecules to enter into or exit from the caldera. 

A study of the nature of trajectories that cross a two dimensional caldera potential was given in \cite{collins2014}.   The caldera potential energy surface studied in that paper possessed a symmetry (to be described shortly), and the effect of asymmetry, or ‘’stretching’’  of the potential, on trajectories was also considered.  In  \cite{katsanikas2018} an analysis of the phase space structures that determined the different behaviors of trajectories was given for the symmetric caldera potential.  In particular, we investigated the mechanisms of trapping trajectories and  of dynamical matching in the symmetrical caldera potential energy surface. The trajectories that have initial conditions on the dividing surfaces of the unstable periodic orbits of the lower saddles are guided from the invariant manifolds of the periodic orbits until they are trapped from the invariant manifolds of the unstable periodic orbits that exist in the central region of the caldera. The trajectories that have initial conditions at the central region of the caldera have two options. The first option is to lie on or are inside the Kolmogorov-Arnold-Moser (KAM) tori that surround the stable periodic orbits of the central area.  The second option is  to be  trapped by the invariant manifolds   of the  unstable  periodic orbits of the central region until they are transported from the unstable invariant manifolds to the exit from the caldera through the four different regions of saddles. The trajectories that have initial conditions on the dividing surfaces of the unstable periodic orbits of the upper saddles are not trapped but,  on the contrary, we have the phenomenon of dynamical matching \cite{katsanikas2018}. Dynamical matching is the behaviour of  trajectories having initial conditions on the periodic orbit dividing surfaces of the upper saddles,  that go straight across the caldera and exit via the opposite lower saddle. We showed that this occurs when there is no interaction of the invariant manifolds of the unstable periodic orbits  of the upper saddles with the central region of the caldera \cite{katsanikas2018}.   This implies that there is no trapping  of these trajectories in the symmetric caldera potential energy surface. 

In this paper we investigate the possibility of a mechanism for trapping of trajectories that have initial conditions on the periodic orbit dividing surfaces of the upper saddles for the stretched version of the caldera  potential in a manner  that does not exist in the symmetric caldera potential energy surface. By ‘’stretching’’ of the potential we mean that  we scale the coordinate x  (see \ref{eq1}) by a parameter $\lambda$ $0<\lambda \leq 1$ and $\lambda \rightarrow 1$. The symmetric  potential is  obtained for  $\lambda=1$. In this situation the saddles move away from the central minimum in the x-direction as the stretching parameter $\lambda$ becomes smaller (see Fig.\ref{equi} and tables 1, 2, 3 and 4). 

We begin our investigation of the stretched potential by first considering if there is a  critical value of $\lambda$ that controls the existence of dynamical matching? The fact that the parameter $\lambda$ plays a role in the dynamical matching phenomenon was evident in the trajectory studies in \cite{collins2014}. However, no explanation of this behavior was given in terms of phase space structure and transport. In this paper we provide such an explanation.
The paper has 5 sections. In section 1 we give an introduction and in the section 2 we describe  the Hamiltonian system with the stretched version of the caldera potential. In section 3 we investigate  the critical value of $\lambda$ for the trapping of trajectories  and in section 4 we study the phase space structure in order to explain the mechanisms of the trapping  of trajectories and their entrance into or exit from the caldera in this case. The summary and conclusions are given  in section 5.

\section{Hamiltonian Model}
\label{sec.1}

We give a brief description of the caldera potential energy surface and Hamiltonian as described in \cite{collins2014}. The caldera potential has a stable equilibrium point at the center, referred to as the central minimum. This potential has an axis of symmetry, the y-axis. We have also the existence of potential walls around the central minimum. On these potential walls we encounter four 1-index saddles (two for lower values of energy, referred to as the lower saddles,  and two for  higher values of energy, referred to as the upper saddles).  In this paper we consider the stretched version of the caldera potential:

\begin{eqnarray}
\label{eq1}
V(x,y)=c_1(y^2+(\lambda x)^2) + c_2y - c_3((\lambda x)^4 + y^4 - 6 (\lambda x)^2 y^2)
\nonumber\\
\end{eqnarray}

\begin{figure}
% \begin{wrapfigure}{l}[0pt]{6.2cm}%%[0pt]{6.2cm}
 \centering
\includegraphics[angle=0,width=5.0cm]{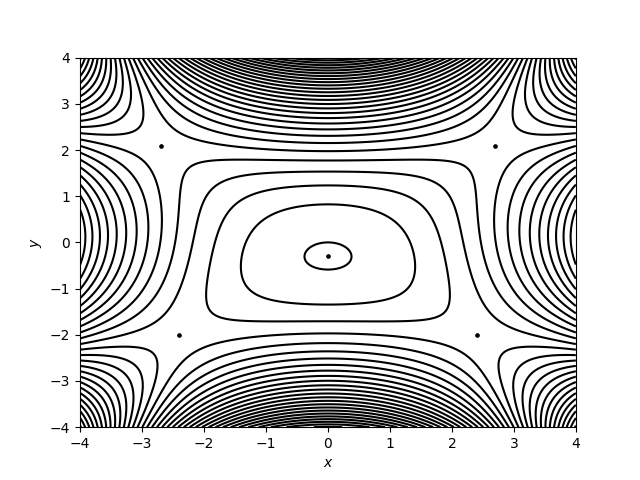}
\includegraphics[angle=0,width=5.0cm]{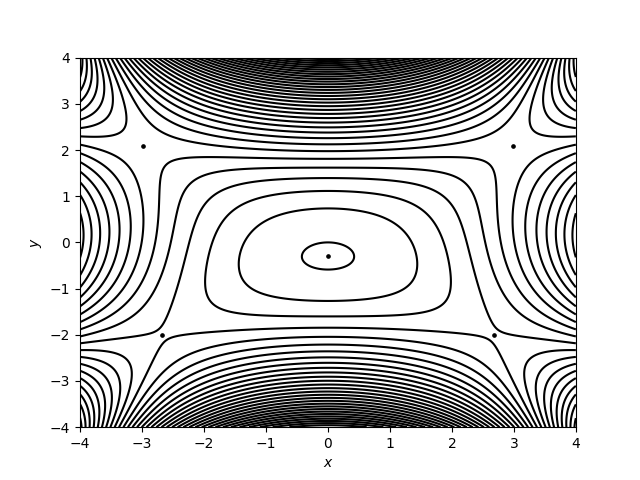}\\
\includegraphics[angle=0,width=5.0cm]{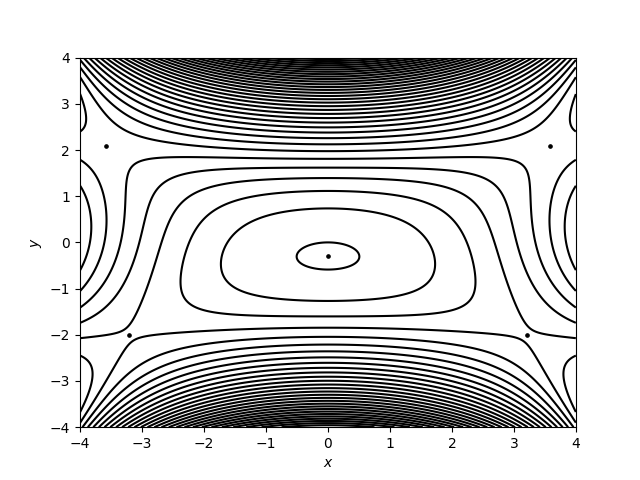}
\includegraphics[angle=0,width=5.0cm]{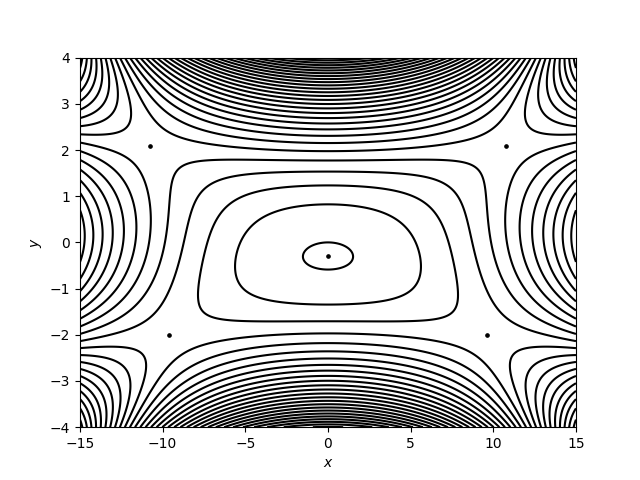}\\
\caption{The stationary points (depicted by black points)  and contours of the potential for $\lambda=0.8$ (upper left panel), $\lambda=0.72$ (upper right panel), $\lambda=0.6$ (lower left panel) and $\lambda=0.2$ (lower right panel).}
\label{equi}
\end{figure}

 \noindent
 The potential parameters are  $c_1=5$, $c_2=3$,$c_3=-3/10$ and $0<\lambda \leq 1$. For $\lambda=1$ we have the symmetric  caldera potential \cite{collins2014,katsanikas2018}. The contours of this potential and the 
 stationary points are depicted for different values of $\lambda$ in  Fig.\ref{equi}. We present the accurate values of the stationary points in the tables 1,2,3 and 4. We can see in these tables and Fig.\ref{equi} that the positions of the saddles move away from the center  of the caldera as the  parameter lambda decreases.

The Hamiltonian of the system is:

\begin{eqnarray}
\label{eq2}
 H(x, y, p_x, p_y )=\frac {p_x^2}{2m}+\frac{p_y^2}{2m} + V(x, y)
\end{eqnarray}

\noindent
with  potential $V(x,y)$  eq. \eqref{eq1} and $m=1$. The numerical value of the hamiltonian we will call it as energy  E.  

The equations of motion are:

\begin{eqnarray}
\label{eq3}
 \dot x= \frac{p_x}{m} \\
 \nonumber\\
 \dot y =\frac{p_y}{m}\\
 \nonumber\\
 \dot p_x=-\frac {\partial V} {\partial x} (x,y)\\
 \nonumber\\
 \dot p_y =-\frac {\partial V} {\partial y} (x,y)
\end{eqnarray}

\begin{table}
\begin{center}
\caption{Stationary points of the potential  given in eq. \eqref{eq1} ("RH" and "LH" are the abbreviations for right hand and left hand respectively) for $\lambda=0.8$}
\begin{tabular}{l  l  l  l}
\hline
Critical point & x & y & E \\
\hline
Central minimum & 0.000 & -0.297 & -0.448 \\
Upper LH saddle  & -2.6862 & 2.0778 & 27.0123 \\
Upper RH saddle  & 2.6862  &  2.0778 & 27.0123 \\
Lower LH saddle & -2.4039 & -2.003  & 14.767  \\
Lower RH saddle & 2.4039  & -2.003 & 14.767 \\
\hline
\end{tabular}
\end{center}
\label{ta08}
\end{table}

\begin{table}
\begin{center}
\caption{Stationary points of the potential  given in eq. \eqref{eq1} ("RH" and "LH" are the abbreviations for right hand and left hand respectively) for $\lambda=0.72$}
\begin{tabular}{l  l  l  l}
\hline
Critical point & x & y & E \\
\hline
Central minimum & 0.000 & -0.297 & -0.448 \\
Upper LH saddle  & -2.9846& 2.0778 & 27.0123 \\
Upper RH saddle  & 2.9846 &  2.0778 & 27.0123 \\
Lower LH saddle & -2.6711 & -2.003  & 14.767  \\
Lower RH saddle & 2.6711& -2.003 & 14.767 \\
\hline
\end{tabular}
\end{center}
\label{ta072}
\end{table}

\begin{table}
\begin{center}
\caption{Stationary points of the potential  given in eq. \eqref{eq1} ("RH" and "LH" are the abbreviations for right hand and left hand respectively) for $\lambda=0.6$}
\begin{tabular}{l  l  l  l}
\hline
Critical point & x & y & E \\
\hline
Central minimum & 0.000  & -0.297 & -0.448 \\
Upper LH saddle  & -3.5815 & 2.0778 & 27.0123 \\
Upper RH saddle  & 3.5815 &  2.0778 & 27.0123 \\
Lower LH saddle & -3.2053 & -2.003  & 14.767  \\
Lower RH saddle & 3.2053 & -2.003 & 14.767 \\
\hline
\end{tabular}
\end{center}
\label{ta06}
\end{table}

\begin{table}
\begin{center}
\caption{Stationary points of the potential  given in eq. \eqref{eq1} ("RH" and "LH" are the abbreviations for right hand and left hand respectively) for $\lambda=0.2$}
\begin{tabular}{l  l  l  l}
\hline
Critical point & x & y & E \\
\hline
Central minimum & 0.000 & -0.297 & -0.448 \\
Upper LH saddle  & -10.7446 & 2.0778 & 27.0123 \\
Upper RH saddle  & 10.7446 &  2.0778 & 27.0123 \\
Lower LH saddle & -9.6158 & -2.003  & 14.767  \\
Lower RH saddle & 9.6158 & -2.003 & 14.767 \\
\hline
\end{tabular}
\end{center}
\label{ta02}
\end{table}

\section{Investigation of the Behaviour of Trajectories as $\lambda$ is Varied}
\label{sec.2}
\begin{figure}
% \begin{wrapfigure}{l}[0pt]{6.2cm}%%[0pt]{6.2cm}
 \centering
\includegraphics[angle=0,width=5.0cm]{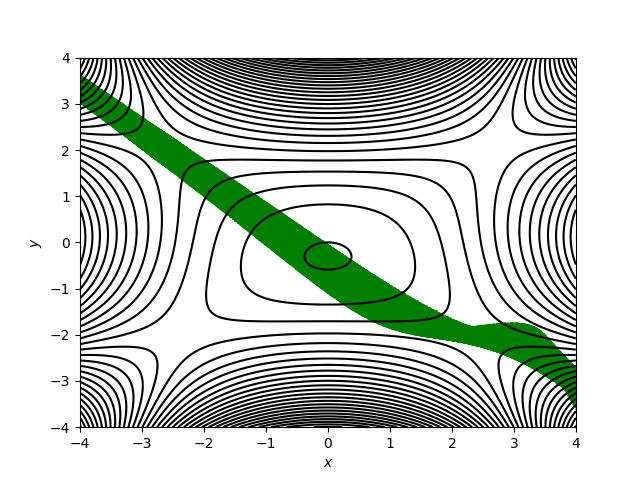}
\includegraphics[angle=0,width=5.0cm]{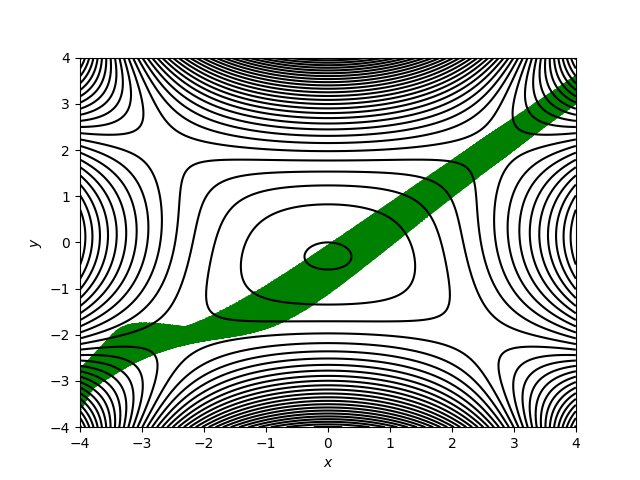}\\
\includegraphics[angle=0,width=5.0cm]{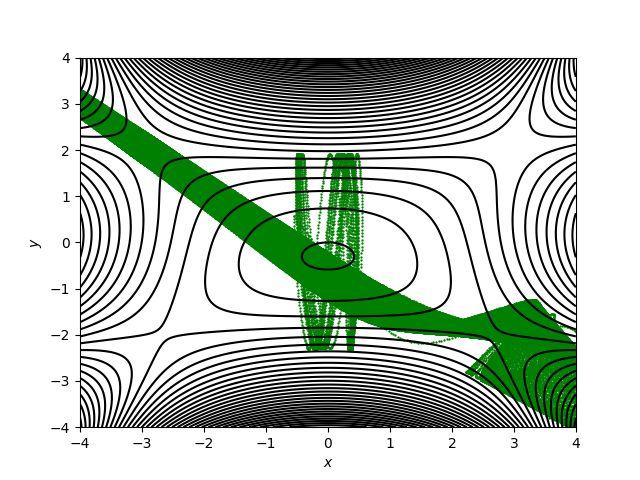}
\includegraphics[angle=0,width=5.0cm]{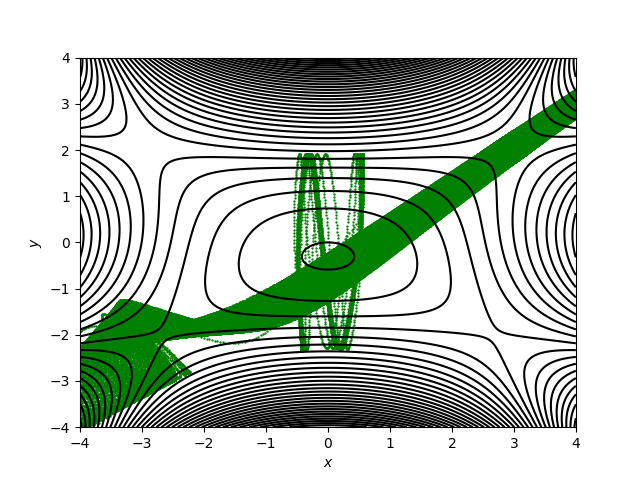}\\
\includegraphics[angle=0,width=5.0cm]{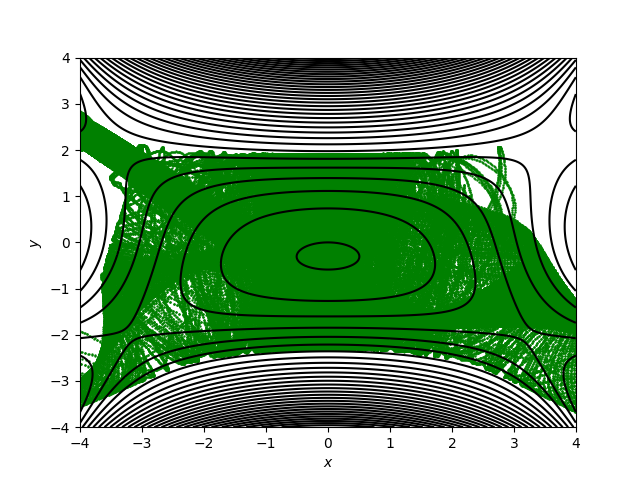}
\includegraphics[angle=0,width=5.0cm]{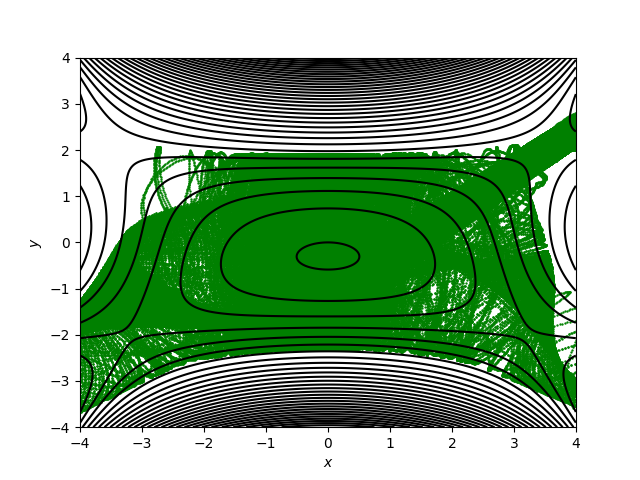}\\
\includegraphics[angle=0,width=5.0cm]{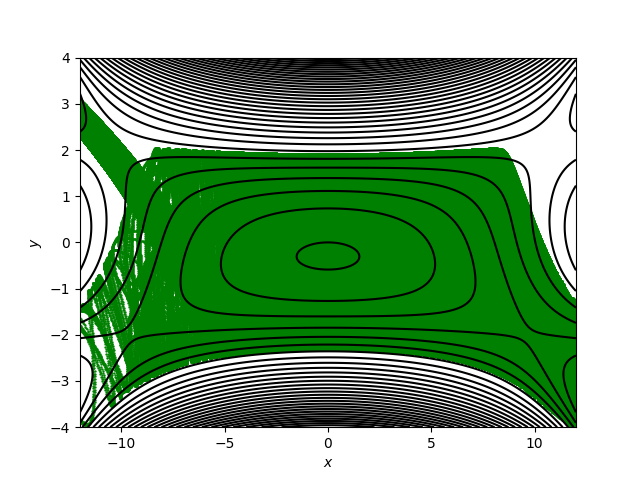}
\includegraphics[angle=0,width=5.0cm]{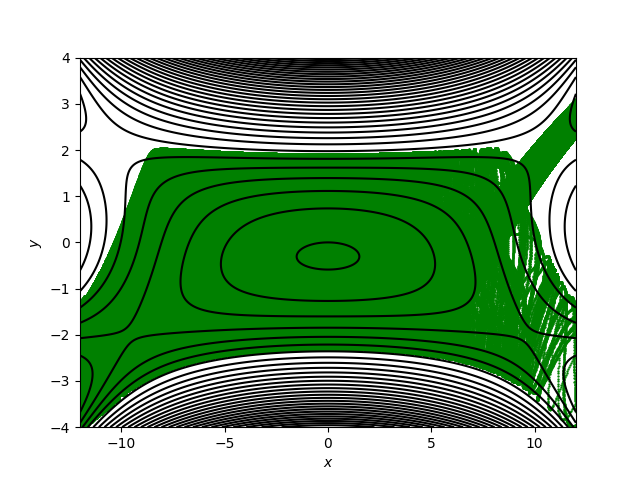}\\
\caption{Trajectories that have initial conditions on the dividing surfaces of the upper left hand saddle and upper right hand saddle, for E=28, for different values of $\lambda$: 
First row of panels: $\lambda =0.8$, second row of panels: $\lambda=0.72$, third row of panels: $\lambda=0.6$ and fourth  row  of panels: $\lambda=0.2$.}
\label{divcc}
\end{figure}

In this section we investigate the  behaviour of trajectories having initial conditions on the dividing surfaces of the  periodic orbits of the upper saddles versus the parameter  $\lambda$. The construction of dividing surfaces associated with periodic orbits for the caldera is described in detail in \cite{katsanikas2018}. The symmetric caldera  problem is studied in  \cite{katsanikas2018} and it is equivalent to the case $\lambda=1$. In the symmetric case the trajectories  that begin from the dividing surfaces of the periodic orbits of upper saddles  are not trapped in the central area of the caldera but they travel straight across the caldera and they exit via the region of the opposite lower saddle (dynamical matching \cite{katsanikas2018}). In this section we  investigate  if  there is a critical value of $\lambda$ for which we have  trapping of trajectories in the central region of the Caldera. 
We decrease the value of $\lambda$ from $1$ and we compute the trajectories, that have initial conditions on the dividing surfaces of the periodic orbits of the upper saddles, for 5 time units (time that we used in  \cite{katsanikas2018}). In  Fig. \ref{divcc} we observe that  trajectories are  trapped for values   $\lambda$  smaller than 0.72. This is the critical value for which we have trapping of trajectories in the central region of the caldera (see for example the cases for   $\lambda=0.72$  $\lambda=0.6$ and  $\lambda=0.2$ in Fig. \ref{divcc}). This implies that we have two cases to consider:

\begin{enumerate}
\item {\textbf First case $0.72< \lambda \leq 1$:} In the first case the trajectories  that have initial conditions on the dividing surfaces  of the periodic orbits of the upper saddles  are not trapped in the central region of the caldera (Dynamical matching). This case is studied in a previous paper for the representative case $\lambda=1$ (symmetric case of the Caldera).  
\item  {\textbf Second case $ \lambda \leq 0.72$:} In this case the trajectories that have initial conditions on the dividing surfaces  of the periodic orbits of upper saddles  are  trapped in the central region of the caldera. This is the case that  is studied in this paper.
\end{enumerate}

\section{Phase space Structure} 
\label{sec.3}

In this section, we study the phase space structures associated with the periodic orbits of the upper saddles in order to understand  the mechanism that is responsible for the trapping in the central area of the caldera for values of $\lambda$ lower that 0.72 (second case). We will study the representative case for $\lambda=0.6$  and for E=28 (the same value of energy that was considered in section \ref{sec.2}). We use the method of the Poincar{\'e}  section (see \cite{katsanikas2018})  to investigate  the phase space structure. As we can see in Fig. \ref{divcc} the trajectories that have initial conditions on the dividing surfaces associated with  the periodic orbits of the upper saddles are trapped for long times before they exit via the regions of the two lower saddles.

\begin{figure}
% \begin{wrapfigure}{l}[0pt]{6.2cm}%%[0pt]{6.2cm}
 \centering
\includegraphics[angle=0,width=12.0cm]{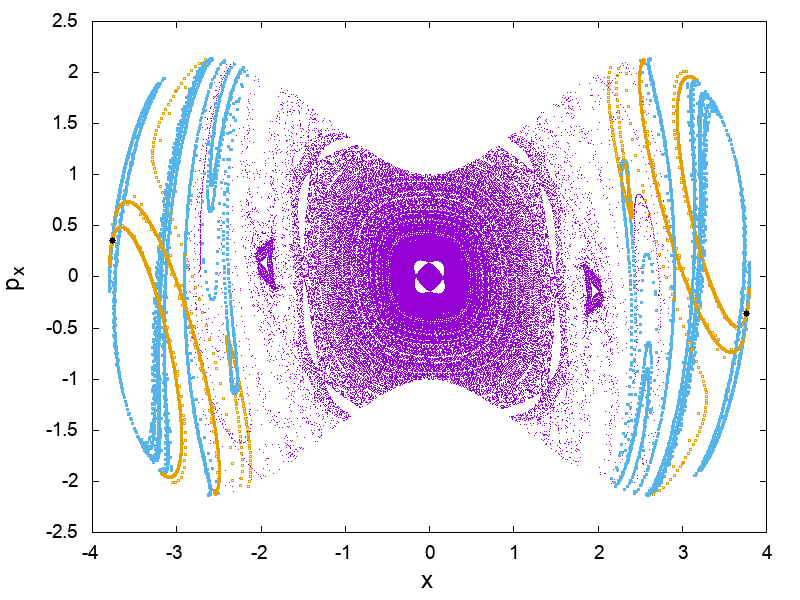}\\
\caption{ The  Poincar{\'e}  section $y=1.894605483735$  with $p_y>0$ for E=28. We discern invariant curves around the stable periodic orbits of the family of the central minimum and  the invariant manifolds (unstable with cyan color - stable with orange color) of the unstable periodic orbits (black points)  of the upper saddles.}
\label{poin1}
\end{figure}

\begin{figure}
% \begin{wrapfigure}{l}[0pt]{6.2cm}%%[0pt]{6.2cm}
 \centering
\includegraphics[angle=0,width=12.0cm]{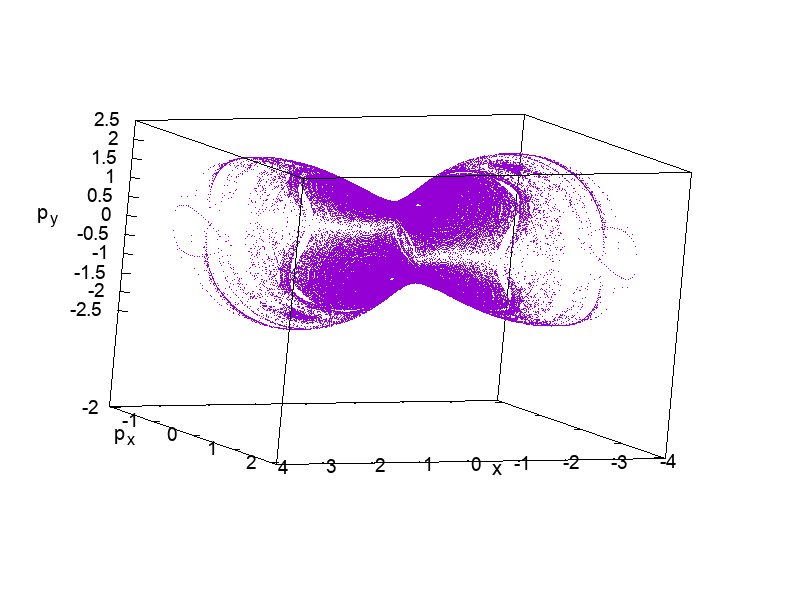}\\
\includegraphics[angle=0,width=12.0cm]{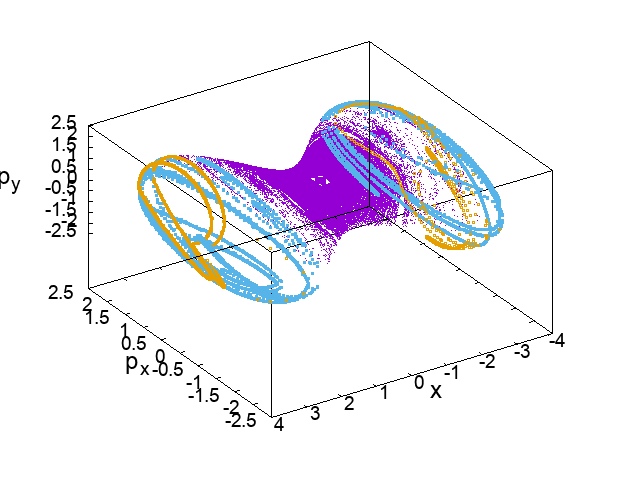}\\
\caption{ The 3D  space $(x,p_x,p_y)$ on the section $y=1.894605483735$ on the 4D phase space  for E=28. We depict also the unstable (with cyan color) and stable (with orange color) invariant manifolds of the periodic orbits of the upper saddles  Upper panel:  with a viewpoint ($79^{o},156^{o}$) and Lower panel: with a viewpoint   
($133^{o},213^{o}$). }
\label{poin3D}
\end{figure}

In Fig. \ref{poin1} we see invariant curves around the central point. These invariant curves are associated with the
stable periodic orbit of the central minimum on the Poincar{\'e}  section  $y=1.894605483735$ with $p_y>0$. Around the central point we see a chain of eight islands. Two upper islands and other two lower islands appear to be cut by the energy surface 
(Fig. \ref{poin1}). This is an artifact of the particular Poincar\'e section chosen which we can see if we plot the $(x,p_x)$  plane of the Poincar{\'e} section with the negative and positive values of $p_y$ we see that the points form a cylindrical surface in the 3D energy surface parametrized by  $(x, p_x, p_y)$ (Fig. \ref{poin3D}) . The islands that seem to be cut in the Fig.\ref{poin1} actually extend from one to the other side of the cylindrical surface (Fig. \ref{poin3D}). Furthermore we see (Fig. \ref{poin1}) many points that appear to stick around the chain of the islands for long times until they leave these structures (stickiness see \cite{contopoulos2002}). This means that there are many trajectories that stick on the invariant tori, that exist in the central area of the caldera, for long time. 

On the left and right side of the figure \ref{poin1}  we observe  the stable (with orange color) and unstable (with cyan color)  invariant manifolds  of the unstable periodic orbits of the upper saddles. The invariant manifolds are  2-dimensional objects in the energy surface of our system and  1-dimensional objects  on the Poincar{\'e} section. The trajectories on manifolds have the capacity to move away from the periodic orbit  on the unstable manifold or to approach the periodic orbit on the stable  manifold. The numerical computation of the unstable and stable  invariant manifolds is based on the integration of many initial conditions (20000 for this paper) in the direction of the unstable eigenvector (the eigenvector that is associated to the real eigenvalue  of the monodromy matrix of the Poincar\'e map  outside the unit circle - see Appendix A in \cite{katsanikas2018})  and of the  stable eigenvector (the eigenvector that is associated to the real eigenvalue  of the monodromy matrix of the Poincar{\'e} map  inside  the unit circle - see Appendix A in \cite{katsanikas2018}) and take their  consequents through the forward (for the case of the unstable manifolds) and backward (for the case of the stable manifolds) integration of the initial conditions.

In Fig. \ref{poin1} we observe that the invariant manifolds begin from the  periodic orbits (that are depicted by black points) and they extend until the central area of the caldera. In addition we see many points are trapped inside the lobes of the invariant manifolds. This means that many trajectories follow the unstable manifolds from the neighbourhood of the unstable periodic orbits to the central area of the caldera. This phenomenon can explain the trapping and the guiding of the trajectories from the periodic orbits of the upper saddles to the central area of the caldera. This is a significant difference between the behaviour of the invariant manifolds in this case and the behaviour of the invariant manifolds in the first case (see \cite{katsanikas2018}). In the first case the invariant manifolds do not interact with the central region of the caldera (see \cite{katsanikas2018}). On the contrary,  in the second case the invariant manifolds are responsible for the trapping and the transport of the trajectories in the central region of the caldera. 

\begin{figure}
% \begin{wrapfigure}{l}[0pt]{6.2cm}%%[0pt]{6.2cm}
 \centering
\includegraphics[angle=0,width=8.0cm]{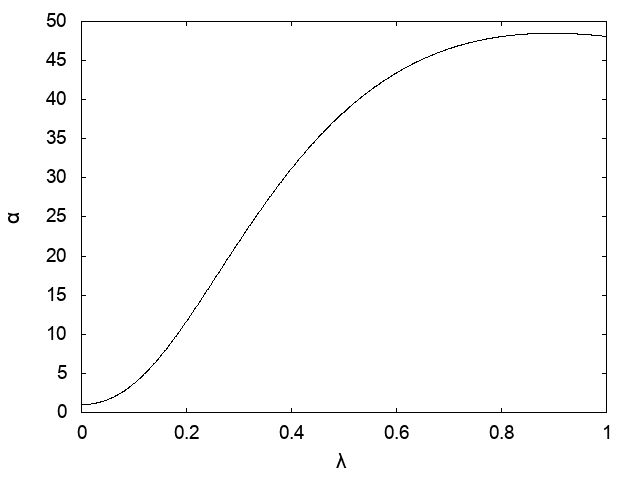}\\

\caption{The stability diagram (the stability parameter $\alpha$ of the periodic orbits versus parameter $\lambda$) of the family of periodic orbits of the right upper saddle.}
\label{stab}
\end{figure}

These results naturally give rise to the following question: Why do the invariant manifolds in the second case interact with the central area of the caldera? In order to answer  this, we have computed (in the Fig. \ref{stab}) the stability diagram of the family of  periodic orbits of the upper right saddle (which, by symmetry,  is identical to the stability diagram for the family of periodic orbits associated with the left upper saddle). The stability diagram describes the evolution of the H{\'e}non stability parameter (see Appendix A in \cite{katsanikas2018}) versus a parameter of the system (in our case the  parameter $\lambda$). In  Fig. \ref{stab} we see  high values of the stability parameter for $\lambda = 1$ and then a small flatness with small fluctuations. For values of $\lambda$  less than 0.8 we have an ongoing reduction in the stability parameter of the unstable periodic orbits. This means that we have a steady decrease in the instability of the periodic orbits, thus increasingly "losing" their ability to repel the orbits in their neighbourhood. This has the consequence that the invariant manifolds begin to transport the trajectories not so far from the area of  the periodic orbits and therefore from the central region of the caldera (since these periodic orbits are located at the upper entrances to the central region of the caldera). Consequently, the invariant manifolds are focussed more towards the central region of the caldera until the value of the stability parameter has diminished sufficiently and then they begin to interact with it (which, as we described, occurs when  λ = 0.72).

The invariant manifolds of  periodic orbit of the  upper right hand  saddle are not connected with the invariant manifolds of the  upper left hand  saddle (see Fig. \ref{poin1}). This can be confirmed if we observe them in the 3D space  $(x, p_x, p_y)$ (Fig. \ref{poin3D}) on the section  $y=1.894605483735$. In this case we see that the invariant manifolds of  one periodic orbit have no relationship with the invariant manifolds of the other periodic orbit as a consequence of the invariant tori between them that divide the space into two regions. This  explains why  trajectories that start from the dividing surfaces of one of the two periodic orbits of the upper saddles do not exit via the region of the other upper saddle. For example if we  see  the first panel of the third row of panels in  Fig. \ref{divcc}  we observe that the trajectories that have initial conditions on the periodic orbit dividing surface of the upper left hand saddle are trapped for long time in the central area of the Caldera and then they exit via the regions of the lower saddles  from the caldera and not via the region of the other upper saddle.

\begin{figure}
% \begin{wrapfigure}{l}[0pt]{6.2cm}%%[0pt]{6.2cm}
 \centering
\includegraphics[angle=0,width=12.0cm]{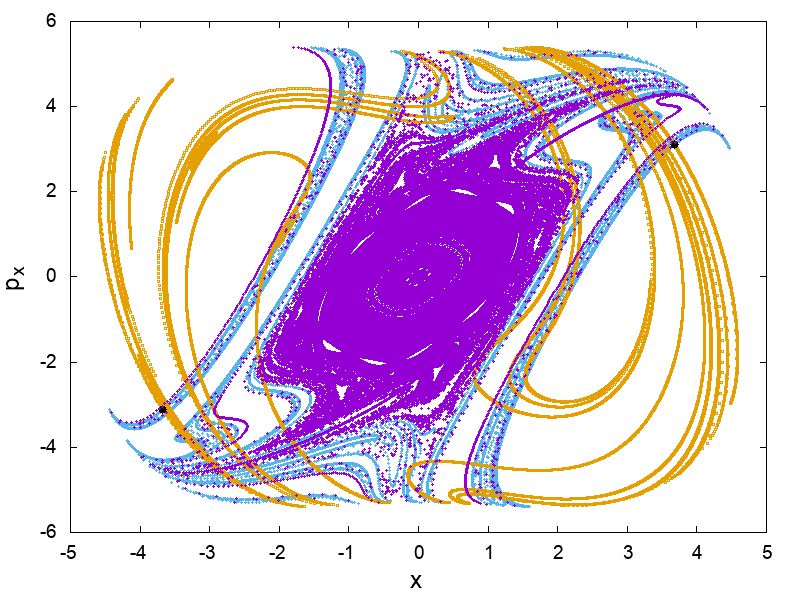}\\
\includegraphics[angle=0,width=12.0cm]{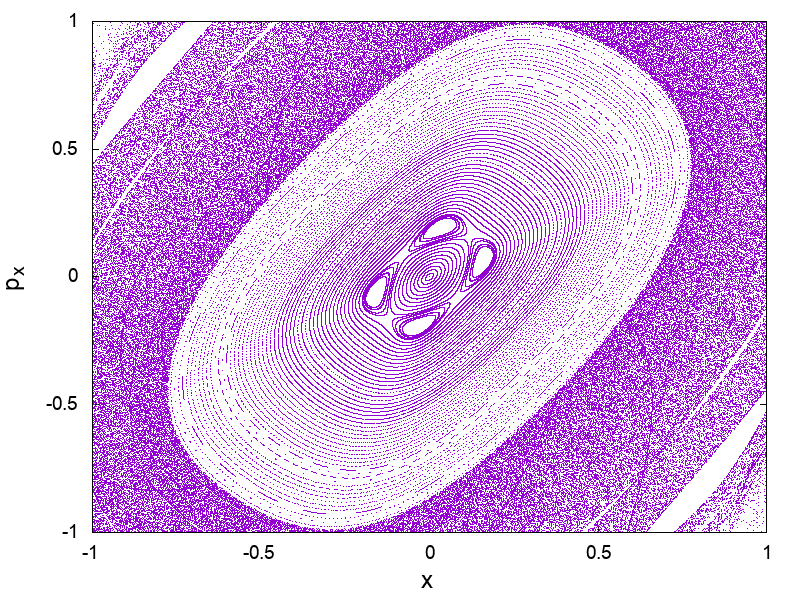}\\
\caption{The  Poincar{\'e}  section $y=-1.8019693$  with $p_y>0$ for E=28. Upper Panel: We observe invariant curves around the stable periodic orbit of the family of the central minimum and  the invariant manifolds (unstable with cyan color - stable with orange color) of the unstable periodic orbits (black points)  of the lower saddles. Lower Panel: An enlargement of the figure of the upper panel in order to see better the   invariant curves around the central point.}
\label{poin2}
\end{figure}

Next we study the phase space structure,using the  Poincar{\'e} section  $y=-1.8019693$ with $p_y>0$,  close to the periodic orbits of the lower saddles in order to understand the mechanism that is responsible for the transport of the trajectories from the central region of the caldera to the exit via the regions of the two lower saddles. In Fig. \ref{poin2} we see invariant curves around the central point that are associated with the stable periodic orbit of the family of the central minimum. This means that there are many trajectories that lie forever on the invariant tori in the central area of the caldera. In addition, we observe that  many points  are  inside the lobes of the  invariant manifolds  of the two periodic orbits of the lower saddles. This means that many trajectories  that are in the central region of the caldera  are trapped by the invariant manifolds and then they follow the stable manifolds  to the neighbourhood of the periodic orbits (the regions of the lower saddles). We also observe that all the available phase space that is outside the stable region around the periodic orbit of the central minimum  is occupied by the invariant manifolds. All the trajectories that are in the central region of the caldera and outside the stable region around the stable periodic orbit of the central minimum can be captured by the invariant manifolds of the periodic orbits of the lower saddles. This can explain why the trajectories that start on the dividing surfaces of the periodic orbits of the  upper saddles exit via the regions of the lower saddles. In addition we observe in the Fig.\ref{poin2} that the invariant manifolds of the lower saddles are not divided in two regions  as in the case of the upper saddles. This is because  the invariant tori do not divide entirely the space between the regions of the two periodic orbits (Fig. \ref{poin2}). Consequently many trajectories, that previously  have been trapped by the invariant manifolds of one periodic orbit, are trapped by the invariant manifolds of the other periodic orbit and they follow the stable manifold of this periodic orbit until they exit the  caldera.

\section{Conclusions}

In previous papers (\cite{collins2014}, \cite{katsanikas2018})   we studied the phenomenon of  dynamical matching exhibited by  trajectories that have initial conditions  on the periodic orbit dividing surfaces associated with the upper saddles for the  case of the symmetric Caldera potential. In this paper we considered the asymmetric caldera potential where the asymmetry  was controlled by a single parameter, $\lambda$, where $\lambda = 1$ corresponds to the symmetric caldera potential energy surface. We found that there is a critical value of the parameter $\lambda$ ($\lambda =0.72$) which for  values of $\lambda$  smaller than this, we have trapping of the trajectories in the central region  of the caldera  and no dynamical matching. This is because the stability parameter of the unstable periodic orbits of the upper saddles decreases for values of $\lambda$ less than $0.8$ and the invariant manifolds are focussing more and more towards the central region of the caldera until they start to interact with it ($\lambda$ = 0.72). The trajectories are trapped  by the invariant manifolds of the unstable periodic orbits of the upper saddles and they are guided from the unstable invariant manifolds into the central region. Then the trajectories can be trapped by the invariant manifolds of the periodic orbits of the lower saddles and they follow their stable manifolds to the neighbourhood of these periodic orbits and then exit from the caldera via the regions of the two lower saddles. In addition, the invariant manifolds of the two periodic orbits of the two upper saddles do not interact with each other because they   are divided into two different regions. This is  because of the existence of topological barriers (invariant tori) between them. A consequence of this is that trajectories having initial conditions on a dividing surface associated with one of two  periodic orbits associated with the upper saddles do not exit via the region of the other upper saddle.

%\clearpage

\nonumsection{Acknowledgments} We acknowledge the support of EPSRC Grant No.~EP/P021123/1 and ONR Grant No.~N00014-01-1-0769.

%\end{multicols}

\end{document}